# Women in Physics in Cyprus: A First Report

## Martha Constantinou[1]

*Department of Physics, University of Cyprus, Nicosia, Cyprus*

marthac@ucy.ac.cy

**Abstract.** This paper reviews the status of women in science, physics in particular, in Cyprus. We describe the development of physics in the country, focusing on the contributions and participation of women. We present statistical data for the last several years, reviewing the percentage of women who are pursuing physics as a subject of study or as a profession. We report the gender ratios at different career stages and find that while women are well represented in undergraduate studies, female physicists are underrepresented in senior positions. We discuss factors that might affect the career evolution of women in physics in Cyprus.

## RESEARCH AND DEVELOPMENT EXPENDITURE

The Cyprus Research Promotion Foundation (RPF) is the main research funding agency in Cyprus. The Foundation has developed several funding programs aimed at implementing high-level research in several thematic areas, developing new products and services for the benefit of Cypriot enterprises, promoting the upgrading of the research infrastructure, developing research collaborations, and utilizing the human research potential of Cyprus. The research and development (R&D) expenditure in Cyprus was last measured at 0.47% of GDP in 2012, according to the World Bank [1]. R&D covers basic research, applied research, and experimental development.

## GENDER EQUALITY ANALYSIS

A number of committees exist in the public sector that work to promote gender equality in their specific sphere of competence. However, many of these committees lack good visibility. In addition, rather than having the support of experts, the committees are formed by individuals in high-ranking government positions in Cyprus who participate in a number of these committees. Unfortunately there is no room for new faces, more qualified people, or alternative voices. In general, the issues with gender policies in Cyprus have to do with active implementation: raising awareness among citizens about new legislation and providing training about gender policies to employers and decision makers [2].

Gender equality is important for every society, but it is even more crucial for Cyprus because talented people are our main resource, and we cannot afford not to develop all the talent that we have. Unless we include women and minorities, we will not be able to reach our full capacity, thus compromising our competitiveness.

## REPRESENTATION OF WOMEN IN PHYSICS

The University of Cyprus is the only institution in the country that offers a BS, MS, and PhD in physics. It accepts, on average, 30 undergraduate students in Physics per year; the number of graduate students drops significantly. The enrollment of females in undergraduate physics programs is higher than that of males; however, the number decreases in graduate programs (see Fig. 1). Unfortunately women are underrepresented in the PhD program, although there are more women than men among the top students.

---

[1] Current electronic address: marthac@temple.edu

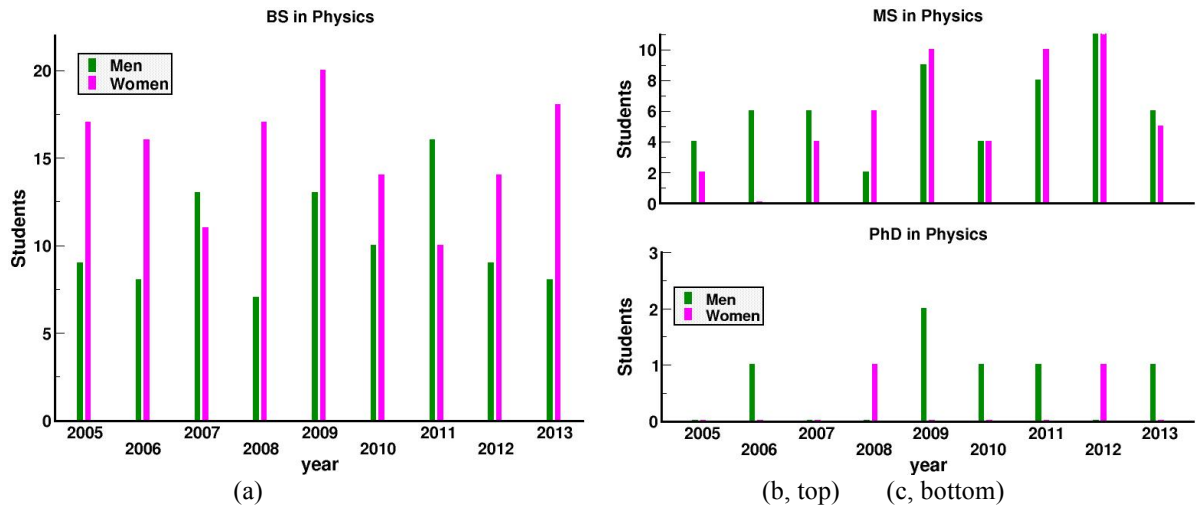

**FIGURE 1.** Statistics for both genders at the University of Cyprus [3] for (a) BS degree in physics,
(b) MS degrees in physics and physics principles, and (c) PhD in physics.

The large number of female Cypriot physicists is not reflected in their participation in scientific activities, nor is it apparent in high-level/faculty positions. Teaching at the high school level is the main professional activity for female Cypriot physicists. In fact, there are more female high school teachers than their male counterparts, which may be due to the security of a permanent position. Table 1 shows the percentage of women in research and academic positions in the Departments of Pure and Applied Sciences at the University of Cyprus. The numbers reveal disproportionate participation of women in high-level research in physics; pursuing a faculty position at universities is much more difficult for women mostly because of societal pressure to focus on family and rather than on research.

**TABLE 1.** Percentage of female staff in the Departments of Pure and Applied Sciences [3].

| Position | Biology | Chemistry | Computer Science | Mathematics | Physics |
|---|---|---|---|---|---|
| Postdoc | 40.0% | 75.0% | 25.0% | 12.5% | 12.5% |
| Faculty | 23.1% | 21.4% | 17.4% | 14.3% | 6.7% |

## SUMMARY AND CONCLUSIONS

There are many talented and intelligent female Cypriot physicists, but this is not reflected by their participation in scientific research and at high positions in universities. We believe that the main reason is the challenge of balancing the high demand of a research career and family obligations. We find it necessary to review our policies, include experts in committees promoting gender equality, and adopt strategies (e.g., government campaigns) successfully applied in other countries with similar issues. We must change the work environment so that women can combine career and family. Science, physics in particular, is not promoted as an option for women in secondary education, which can affect a student's future career. There is a lack of science-related extracurricular activities, and the infrastructure (e.g. laboratories, supercomputing centers) is insufficient. It is also very important to motivate female undergraduate students to continue with postgraduate studies. One way to do so is to provide financial support for female students to attend summer schools, workshops, and conferences, and most importantly to participate in research projects. Last but not least, we believe that it is important to establish a platform for female physicists to discuss and exchange ideas**.**